\documentclass[aps,pra,groupaddress,showpacs,twocolumn]{revtex4-1}
\usepackage{graphicx,bm,amsmath,amsfonts,bbold, amssymb,color,microtype}

\newcommand{\be}{\begin{equation}}
\newcommand{\ee}{\end{equation}}
\newcommand{\bea}{\begin{eqnarray}}
\newcommand{\eea}{\end{eqnarray}}
\newcommand{\im}[1]{\mbox{Im}[#1]}
\newcommand{\re}[1]{\mbox{Re}[#1]}

\begin{document}

\title{Scattering in $\cal PT$ and $\cal RT$ Symmetric Multimode Waveguides: Generalized Conservation Laws and Spontaneous Symmetry Breaking beyond One Dimension}

\author{Li Ge}
\email{li.ge@csi.cuny.edu}
\affiliation{\textls[-18]{Department of Engineering Science and Physics, College of Staten Island, CUNY, Staten Island, NY 10314, USA}}
\affiliation{The Graduate Center, CUNY, New York, NY 10016, USA}
\author{Konstantinos G. Makris}
\affiliation{Crete Center for Quantum Complexity and Nanotechnology, Department of Physics, University of Crete, P.O. Box 2208, 71003, Heraklion, Greece}
\author{D. N. Christodoulides}
\affiliation{College of Optics and Photonics -- CREOL, University of Central Florida, Orlando, Florida 32816, USA}
\author{Liang Feng}
\affiliation{Department of Electrical Engineering, The State University of New York at Buffalo, Buffalo, NY 14260, USA}

\begin{abstract}
We extend the generalize conservation law of light propagating in a one-dimensional $\cal PT$-symmetric system, i.e., $|T-1|=\sqrt{R_LR_R}$ for the transmittance $T$ and the reflectance $R_{L,R}$ from the left and right,  to a multimode waveguide with either $\cal PT$ or $\cal RT$ symmetry, in which higher dimensional investigations are necessary. These conservation laws exist not only in a matrix form for the transmission and reflection matrices; they also exist in a scalar form for real-valued quantities by defining generalized transmittance and reflectance. We then discuss, for the first time, how a multimode $\cal PT$-symmetric waveguide can be used to observe spontaneous symmetry breaking of the scattering matrix, which typically requires tuning the non-hermiticity of the system (i.e. the strength of gain and loss). Here the advantage of using a multimode waveguide is the elimination of tuning any system parameters: the transverse mode order $m$ plays the role of the symmetry breaking parameter, and one observes the symmetry breaking by simply performing scattering experiment in each waveguide channel at a single frequency and fixed strength of gain and loss.
\end{abstract}


\maketitle

\section{introduction}

Parity-Time ($\cal{PT}$) symmetric optical systems have attracted growing interest in the past several years \cite{El-Ganainy_OL06,Moiseyev,Musslimani_prl08,Makris_prl08,Kottos,Guo,mostafazadeh,Makris_PRA10,Longhi,CPALaser,conservation,Robin,EP9,Ruter,Lin,Feng,Feng_NM,Feng2,Walk,Hodaei,Yang}. These systems are non-hermitian due to the presence of gain and loss, which are delicately balanced to make the refractive index satisfying $n(x)=n^*(-x)$ with respect to a chosen symmetry plane at $x=0$. The plethora of findings in such systems are tied to the spontaneous symmetry breaking at an exceptional point (EP) \cite{EP1,EP2,EPMVB,EP3,EP4,EP5,EP6,EP7,EP8}. This spontaneous symmetry breaking was first suggested in non-hermitian quantum mechanism \cite{Bender1,Bender2,Bender3} and later found in the evolution of waves in the paraxial regime \cite{El-Ganainy_OL06,Moiseyev,Musslimani_prl08,Makris_prl08}, which takes the system from a regime of real energy eigenvalues to complex conjugate pairs of eigenvalues.

Recently it was found that the scattering eigenstates of a $\cal PT$-symmetric system also display a spontaneous symmetry breaking \cite{CPALaser}, independent of its shape and dimension: the eigenvalues of the scattering ($S$) matrix can remain on the unit circle in the complex plane, conserving optical flux despite the non-hermiticity; the symmetry breaking results in pairs of scattering eigenvalues with inverse moduli \cite{CPALaser,conservation,Robin}. Using the symmetry property of the $S$ matrix \cite{CPALaser} or equivalently that of the transfer matrix \cite{Longhi}, one can derive a generalized conservation law in one dimension (1D) \cite{conservation}, i.e.,
\be
|T-1|=\sqrt{R_LR_R},\label{eq:law1D}
\ee
in contrast to the usual conservation relation $T+R_{L(R)}=1$ when the system is hermitian. Here $T$ is the transmittance from either the left or right side (they are identical due to optical reciprocity \cite{Haus,Collin,Landau}) and $R_{L,R}$ are the reflectance from the two sides, respectively. At an accidental reflection degeneracy where $R_L=R_R\equiv R$, the generalized conservation law above indicates two possible scenarios, where either $T+R=1$ or $T-R=1$. The former is identical to the hermitian conservation law even though the system is non-hermitian, and the transmittance in the latter is clearly super-unitary.

\begin{figure}[t]
\begin{center}
\includegraphics[width=\linewidth]{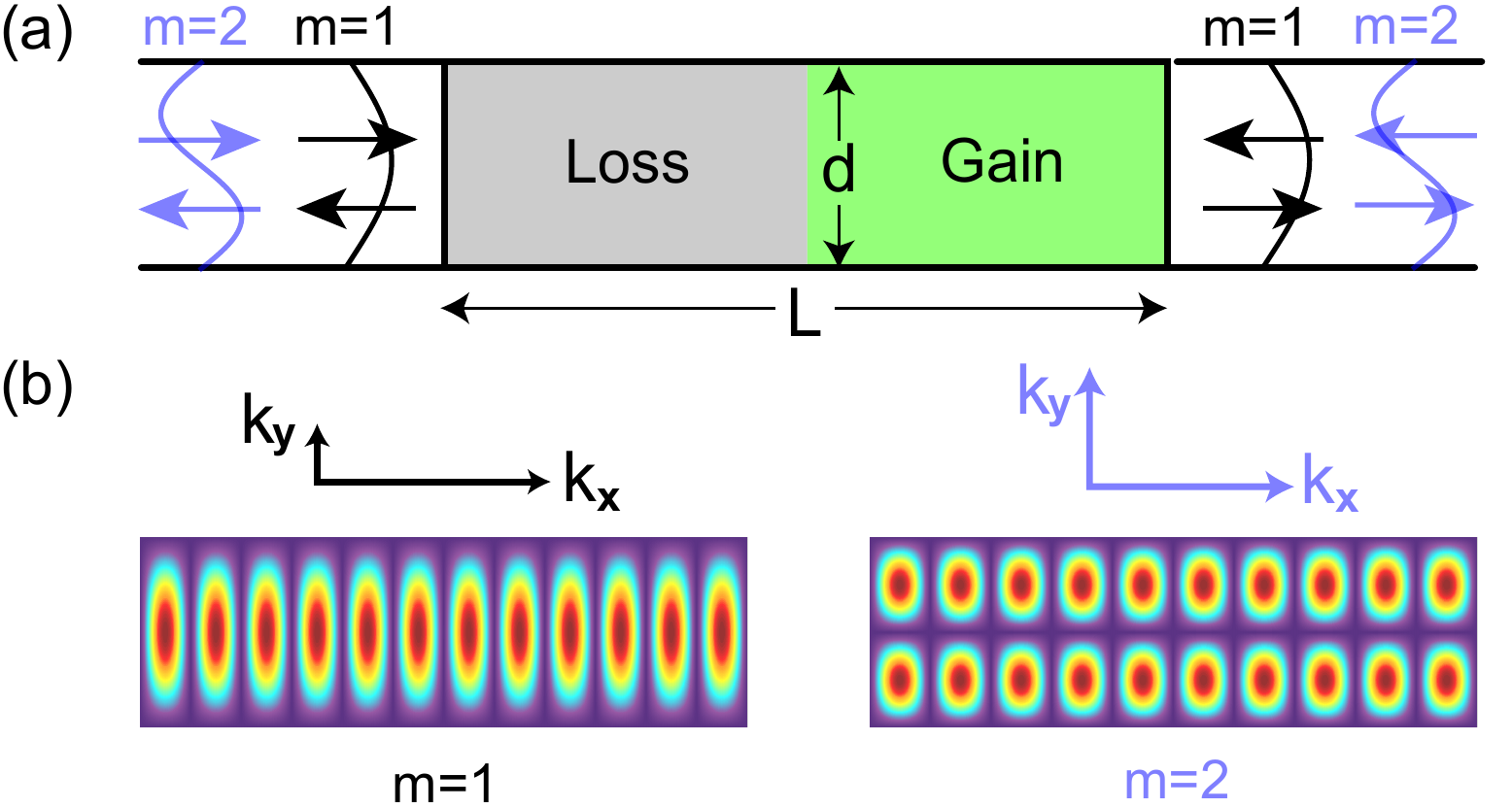}
\caption{(Color online) (a) Schematic of a $\cal PT$-symmetric multimode waveguide of length $L$ and width $d$. Incoming and outgoing channels of transverse order $m=1,2$ are illustrated in both the left and right ``leads" of the waveguide. (b) Standing wave patterns with transverse order $m=1,2$ in the leads. The longitudinal and transverse wavevectors corresponding to their right-traveling components are contrasted schematically.}
\label{fig:schematic}
\end{center}
\end{figure}

In this report we first extend the generalize conservation law above in 1D to higher dimensions, i.e., in a multimode waveguide with either $\cal PT$ or Rotation-Time ($\cal RT$) symmetry \cite{EP9} (see Fig.~\ref{fig:schematic}), where $\cal R$ is rotation by $\pi$ about a given axis. Using the matrix representation of $\cal P$ and $\cal R$ in block form, we show that a similar expression exists for the transmission and reflection matrices, which holds independent of the details of the channels, i.e., whether they are sinusoildal waves in a ridge waveguide or cylindrical waves in an optical fiber. By further defining the generalized transmittance and reflectance suitable for $\cal PT$- and $\cal RT$-symmetric systems, we also reduce these conservation laws to their scalar forms with only real-valued quantities.

We then discuss how a multimode $\cal PT$-symmetric waveguide can be used to observe spontaneous symmetry breaking of the $S$ matrix. It was suggested in Ref.~\cite{CPALaser} that this symmetry breaking can be observed in 1D (i.e., in a single-mode waveguide) by tuning either the gain and loss strength $\tau$ of the system or the product of the waveguide length $L$ and the frequency $\omega$ of the incident light. Such an approach was successfully implemented in phononics, where clear regimes of $\cal PT$-symmetric and $\cal PT$-broken phases were observed \cite{Pagneux}. In the optical regime however, this approach was not well received due to the stringent requirement on maintaining $\cal PT$ symmetry while tuning across the symmetry breaking point. In our current proposal of using a multimode waveguide, the advantage is the elimination of tuning any system parameters; the transverse mode order $m=1,2,\ldots,N$ plays the role of the symmetry breaking parameter, as we will show below. Hence one can observe the symmetry breaking by simply performing scattering experiment in each waveguide channel, at a \textit{single} frequency and \textit{fixed} strength of gain and loss.

\section{Generalized Conservation Laws}

There are two complexities when extending the the generalized conservation law given by Eq.~(\ref{eq:law1D}) to a multimode waveguide. The first one is that Eq.~(\ref{eq:law1D}) only includes three quantities, which are the reflectance from both left and right sides and the transmittance. For a multimode waveguide with $N$ channels however, each of these quantities become a $N\times N$ matrix (unless there is no coupling between the channels in the system), and the increased degrees of freedom prompt us to look for a conservation law in a \textit{matrix} form first. The second complexity is that the three quantities in Eq.~(\ref{eq:law1D}) are non-negative real numbers, with the phases of the transmission coefficient $t$ and reflection coefficients $r_{L,R}$ eliminated through the definition $T=|t|^2$ and $R_{L,R} = |r_{L,R}|^2$. In a multimode waveguide however, it seems difficult to eliminate the phases of all the transmission and reflection coefficients. Therefore, the generalized conservation law is hence more likely an matrix identity for the complex-valued transmission matrix $\bm{t}$ and reflection matrices $\bm{r}_{L,R}$, instead of for the real-valued transmittance matrix and reflectance matrices. Nevertheless, we indeed find a \textit{scalar} and \textit{real-valued} form of the generalized conservation law for both $\cal PT$-symmetric and $\cal RT$-symmetric systems, using generalized definitions of transmittance and reflectance as we show below.

By denoting the amplitudes of the $N$ incoming (outgoing) channels in the left lead by $\bm{\varphi}_\text{in}^L$ ($\bm{\varphi}_\text{out}^L$) and those in the right lead by $\bm{\varphi}_\text{in}^R$ ($\bm{\varphi}_\text{out}^R$), $\bm{t}$ and $\bm{r}_{L,R}$ are defined by
\be
\bm{\varphi}_\text{out}^L = \bm{t}\bm{\varphi}_\text{in}^R, \quad \bm{\varphi}_\text{out}^R = \bm{r}_R\bm{\varphi}_\text{in}^R \label{eq:Rin}
\ee
when there are only incident waves in the right lead, and
\be
\bm{\varphi}_\text{out}^R = \bm{t}^T\bm{\varphi}_\text{in}^L, \quad \bm{\varphi}_\text{out}^L = \bm{r}_L\bm{\varphi}_\text{in}^L \label{eq:Lin}
\ee
when there are only incident waves in the left lead. The superscript ``$T$'' denotes matrix transpose in Eq.~(\ref{eq:Lin}), and it appears due to optical reciprocity \cite{Haus,Collin,Landau}: the transmission coefficient from channel $m$ on the left to channel $m'$ on the right is the same as the reversed process. This is also the case in 1D, where $\bm{t}$ reduces to a complex number and the matrix transpose can be omitted. Here we do not consider systems that break optical reciprocity, including but not limited to magneto-optical systems. 
The $S$ matrix is then defined by
\be
\begin{pmatrix}
\bm{\varphi}_\text{out}^L \\
\bm{\varphi}_\text{out}^R
\end{pmatrix}
= \begin{pmatrix}
\bm{r}_L & \bm{t}\\
\bm{t}^T & \bm{r}_R
\end{pmatrix}
\begin{pmatrix}
\bm{\varphi}_\text{in}^L \\
\bm{\varphi}_\text{in}^R
\end{pmatrix}
\equiv
\bm{S}
\begin{pmatrix}
\bm{\varphi}_\text{in}^L \\
\bm{\varphi}_\text{in}^R
\end{pmatrix},\label{eq:S2D}
\ee
which is a $2N\times 2N$ matrix. Note that we do not include the channels of evanescent waves, which decays exponentially away from the waveguide in both leads.

The derivation of Eq.~(\ref{eq:law1D}) in Ref.~\cite{conservation} is based on the symmetry relation of the transfer matrix \cite{Longhi}. Here we use the symmetry relation of the $S$ matrix instead \cite{CPALaser}, i.e.,
\be
{\cal PT}\bm{S}{\cal PT}=\bm{S}^{-1} \label{eq:PTSPT}
\ee
which is equivalent but more convenient in the case of a $\cal PT$-symmetric multimode waveguide. Here the time-reversal operator $\cal T$ is simply complex conjugate and denoted by a superscript ``$*$'', and the parity operator $\cal P$ along the longitudinal direction can be represented by a matrix permutation $\bm{P}$ satisfying $\bm{P}^2=\bm{1}_{2N}$ \cite{CPALaser}, where $\bm{1}_{2N}$ is the identity matrix of rank $2N$. Therefore, we can rewrite Eq.~(\ref{eq:PTSPT}) as
\be
\bm{P}\bm{S}^*\bm{P}=\bm{S}^{-1}. \label{eq:PTSPT2}
\ee
It is important to note that in a multimode waveguide, the incoming/outgoing channel $m$ on the left side becomes the same incoming/outgoing channel $m$ on the right side after the parity operation, since the parity operation is only about the longitudinal coordinate and leaves the transverse coordinate(s) unchanged. (This is not the case in a $\cal RT$-symmetric waveguide as we shall see below.) Therefore, the parity operator here does not change the order of the left channels or the right channels, and its matrix representation $\bm{P}$ takes the following block form
\be
\bm{P} = \begin{pmatrix}
0 & \bm{1}_N \\
\bm{1}_N & 0
\end{pmatrix}.
\ee
This form of $\bm{P}$ greatly simplifies Eq.~(\ref{eq:PTSPT2}), the left hand side of which becomes
\be
\bm{P}\bm{S}^*\bm{P}
= \begin{pmatrix}
\bm{r}_R & \bm{t}^T\\
\bm{t} & \bm{r}_L
\end{pmatrix}^*.
\ee
Using matrix inversion in block form, we find the following four relations
\begin{gather}
(\bm{r}_R^*)^{-1} = \bm{r}_L - \bm{t}(\bm{r}_R)^{-1}\bm{t}^T, \label{eq:Block1}\\
(\bm{r}_L^*)^{-1} = \bm{r}_R - \bm{t}^T(\bm{r}_L)^{-1}\bm{t}, \label{eq:Block2}\\
\bm{r}_R\bm{t}^* + \bm{t}^T\bm{r}_R^* = 0, \label{eq:Block3}\\
\bm{r}_L^*\bm{t}^T + \bm{t}^*\bm{r}_L = 0, \label{eq:Block4}
\end{gather}
by equating the four $N\times N$ blocks on the left and right sides of Eq.~(\ref{eq:PTSPT2}). Next we combine Eqs.~(\ref{eq:Block1}),(\ref{eq:Block3}) and Eqs.~(\ref{eq:Block2}),(\ref{eq:Block4}) respectively, which lead to
\begin{align}
\bm{t}\bm{t}^*-\bm{1}_N&=-\bm{r}_L\bm{r}_R^*,\label{eq:law2D1}\\
\bm{t}^\dagger\bm{t}^T-\bm{1}_N&=-\bm{r}_R^*\bm{r}_L,\label{eq:law2D2}
\end{align}
where the superscript ``$\dagger$" denotes hermitian conjugate as usual.

Eq.~(\ref{eq:law2D2}) is in fact identical to Eq.~(\ref{eq:law2D1}), once we take its transpose and use the property that $\bm{r}_{L,R}$ are symmetric, which comes from optical reciprocity \cite{Haus,Collin,Landau}, i.e., the scattering coefficient from incoming channel $m$ to outgoing channel $m'$ on the same side is the same as that of the reversed process, for both left- and right-side incidence. In addition, we note that using the properties of matrix trace, i.e., $Tr(\bm{A}^*) = (Tr \bm{A})^*$ and $Tr(\bm{AB}) = Tr(\bm{BA})$ for two arbitrary square matrices $\bm{A}$ and $\bm{B}$, we find that $Tr(\bm{t}\bm{t}^*)\equiv \bar{T}$ is real (see further discussion in the Appendix), which implies that $Tr(\bm{r}_L\bm{r}_R^*)\equiv\bar{R}$ is also real and
\be
\bar{T}+\bar{R}=N \label{eq:law2D3}
\ee
by considering Eq.~(\ref{eq:law2D1}). We will refer to $\bar{T}$ and $\bar{R}$ as the generalized transmittance and reflectance in $\cal PT$-symmetric multimode waveguides.

The corresponding relations to Eqs.~(\ref{eq:law2D1}-\ref{eq:law2D3}) in a hermitian system are given by
\begin{align}
\bm{t}\bm{t}^\dagger - \bm{1}_N &= -\bm{r}_L\bm{r}_L^\dagger, \label{eq:law2D1_h}\\
\bm{t}^\dagger\bm{t} - \bm{1}_N &= -\bm{r}_R^\dagger\bm{r}_R, \label{eq:law2D2_h} \\
T+R_L=T&+R_R=N.\label{eq:law2D3_h}
\end{align}
The first two are derived from $\bm{S}\bm{S}^\dagger=1$, and their traces lead to Eq.~(\ref{eq:law2D3_h}), where $T\equiv Tr(\bm{t}\bm{t}^\dagger)$ and $R_{L,R}\equiv Tr(\bm{r}_{L,R}\bm{r}^\dagger_{L,R})$ are both real \cite{Beenakker}. 
%$T$ is particular important in mesoscopic transport, which gives the electron conductance when multiplied by a system independent constant.

We will refer to Eqs.~(\ref{eq:law2D1}) and (\ref{eq:law2D3}) as the generalized conservation laws in a $\cal PT$-symmetric multimode waveguide. Note that they are valid in general and the details of the waveguide channels are not specified in its derivation, which can be, for example, sinusoidal waves in a ridge waveguide or cylindrical waves in an optical fiber. Therefore, the generalized conservation laws given by Eqs.~(\ref{eq:law2D1}) and (\ref{eq:law2D3}) hold independent of the transverse geometry of the waveguide, as long as the waveguide is $\cal PT$-symmetric.

Eq.~(\ref{eq:law2D1}) reduces to its 1D form given by Eq.~(\ref{eq:law1D}) once we replace $\bm{t}, \bm{r}_{L,R}$ by their scalar form $t,r_{L,R}$ in 1D and take the absolute value of both its left and right sides. In the special case that the waveguide channels do not couple (which occurs, for example, when the transverse and longitudinal coordinates are separable), $\bm{t},\bm{r}_{L,R}$ become diagonal matrices, and we again recover Eq.~(\ref{eq:law1D}) for each waveguide channel, which acts as an independent 1D $\cal PT$-symmetric system. To exemplify how Eq.~(\ref{eq:law1D}) breaks down for the diagonal elements of Eq.~(\ref{eq:law2D1}) in the general case, below we study the scattering of transverse electric (TE) waves in a two-dimensional (2D) waveguide and adopt Dirichlet boundary condition on the sidewalls in the transverse ($y$) direction. Each channel can be labelled by a transverse mode number $m=1,2,\ldots,N$, and the transverse mode profiles are given by $\psi(y)=\sin(m\pi(y/d+1/2))~(y\in[-d/2,d/2])$. Here $d$ is the width of the waveguide, and we have chosen the center axis of the waveguide as the origin of the $y$-axis and assumed $n=1$ in the leads.

\begin{figure}[t]
\begin{center}
\includegraphics[width=\linewidth]{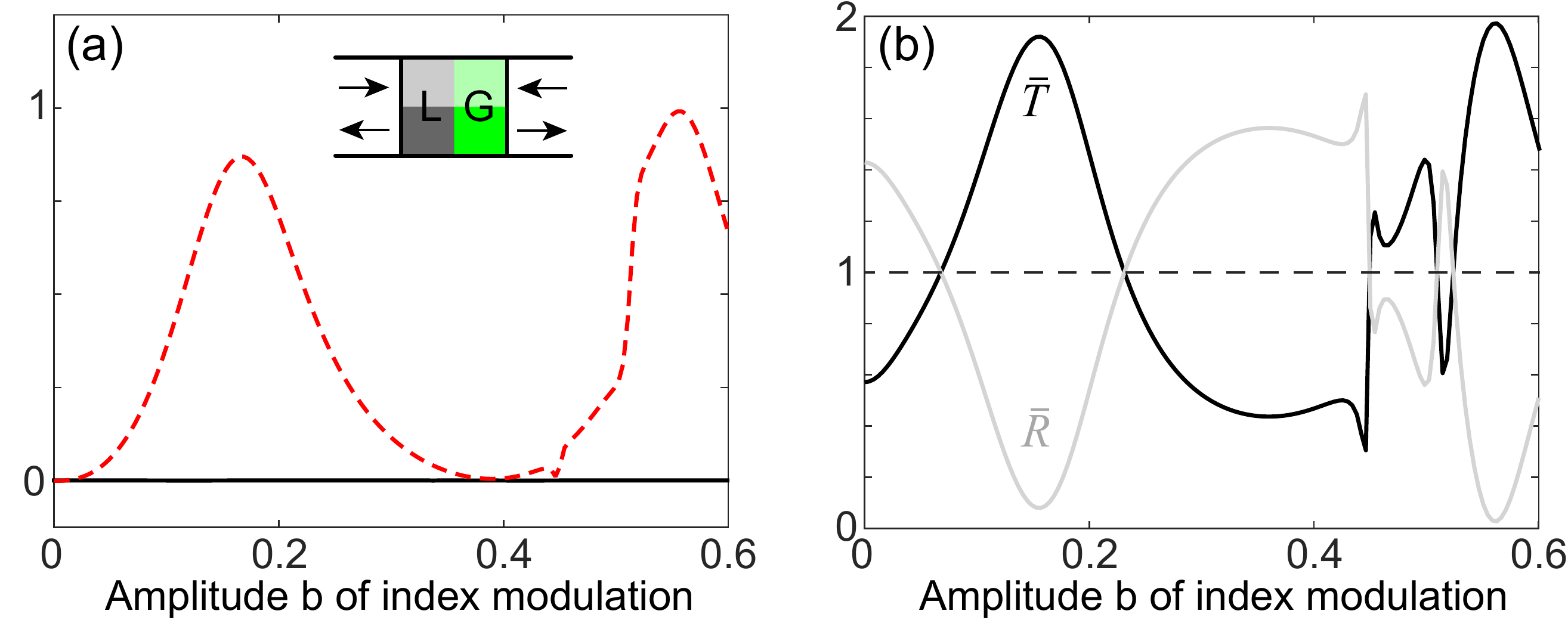}
\caption{Verification of generalized conservation laws in a 2D $\cal PT$-symmetric waveguide with two channels in both the left and right leads. (a) Solid line plots the norm of the difference between the two sides of Eq.~(\ref{eq:law2D1}), confirming the 2D conservation relation (\ref{eq:law2D1}). Dashed line plots the absolute value of the difference of the two sides of Eq.~(\ref{eq:law1D}) for the $m=2$ channel, showing the breakdown of the 1D conservation relation (\ref{eq:law1D}). Inset: the length $L$ and width $d$ of the waveguide are both 2 $\mu$m, and the wavelength is 1550 nm. (b) Black and grey lines plot $\bar{T}$ and $\bar{R}$. Their mirror symmetry about 1 (dashed horizontal line) verifies the 2D scalar conservation relation (\ref{eq:law2D3}), where $N=2$.}
\label{fig:law2D}
\end{center}
\end{figure}

Below we will refer to a system with separable longitudinal and transverse coordinates simply as a separable system. We start with such a separable system of $N=2$ (i.e., two channels for incoming/outgoing waves in both the left and right leads) as illustrated in Fig.~\ref{fig:schematic}(a), which has a uniform refractive index $n=3$ and length $L=2\,\mu$m before gain and loss represented by $\im{n}=\mp0.005$ are introduced to its two halves. We then increase the amplitude $b$ of a stepwise index modulation in the transverse direction, $\Delta n(y)=b\, (y<0);\, -b\, (y>0)$, which introduces and changes the coupling between the two waveguide channels and the system becomes non-separable. As Fig.~\ref{fig:law2D} shows, the generalized conservation law (\ref{eq:law2D1}) holds independent of $b$, while Eq.~(\ref{eq:law1D}) for a single channel no longer holds at a non-zero $b$ and its behavior is correlated with the resonances of the system. %Note that to eliminate the effects of evanescent waves, the left and right ends of the scattering region are chosen to be 3 $\mu$m away from the actual end surfaces of the waveguide in the leads.

% In addition to the generalized conservation law (\ref{eq:law2D1}), Eqs.~(\ref{eq:Block3}) and (\ref{eq:Block4}) provide additional information about the phases of $\bm{t}$ and $\bm{r}_{L,R}$; their 1D correspondences indicate that $r_L$ and $r_R$ are always $\pi/2$ out of phase with $t$ \cite{conservation}, which in turn implies that $r_L$ and $r_R$ themselves are in phase or $\pi$ out of phase; the 1D correspondence of Eq.~(\ref{eq:law2D1}), i.e., $|t|^2-1=-r_Lr_R^*$ could not possibly hold without this relation.

Next we turn to an $\cal RT$-symmetric multimode waveguide. It is straightforward to obtain the symmetry relation of the $S$ matrix,
\be
{\cal RT}\bm{S}{\cal RT} = \bm{S}^{-1},\label{eq:RTSRT}
\ee
and its matrix representation,
\be
\bm{R}\bm{S}^*\bm{R} = \bm{S}^{-1},
\ee
which are similar to Eqs.~(\ref{eq:PTSPT}) and (\ref{eq:PTSPT2}) in the $\cal PT$-symmetric case.
As we have mentioned, the $\pi$-rotation operator $\cal R$ does not necessarily transform the incoming/outgoing channel $m$ on the left to the same incoming/outgoing channel $m$ on the right. Take the 2D waveguide shown in Fig.~\ref{fig:schematic}(a) for example, if we choose the channel wave functions in the left lead to be the same as before, i.e., $\psi(y)=\sin(m\pi(y/d+1/2))$, then they become $(-)^{m+1}\sin(m\pi(y/d+1/2))$ in the right leads because $\cal R$ transforms $y\rightarrow-y$ in addition to $x\rightarrow-x$. In other words, a negative sign is introduced for the even-$m$ channels (whose mode profiles are odd functions of $y$), and $\bm{R}$ takes the following form
\be
%\hspace{-1mm}
\bm{R} = \begin{pmatrix}
0 & \tilde{\bm{1}}_N \\
\tilde{\bm{1}}_N & 0
\end{pmatrix},\,\,
\tilde{\bm{1}}_N\equiv
\begin{pmatrix}
1 & & &0\\
& -1 & &\\
& & \ddots & \\
0& & & (-1)^{N+1}
\end{pmatrix}.
\ee
We note that $\bm{R}$ is symmetric but not symplectic (it satisfies $\bm{R}^T\bm{\Omega}\bm{R}=-\bm{\Omega}$ instead of $\bm{R}^T\bm{\Omega}\bm{R}=\bm{\Omega}$, where $\bm{\Omega}=\left(\begin{smallmatrix}
0 & \bm{1}_N \\
-\bm{1}_N & 0
\end{smallmatrix}\right)$). Using the above block form of $\cal R$, we find that the generalized conservation law in an $\cal RT$-symmetric waveguide is
\be
\bm{t}\tilde{\bm{1}}_N\bm{t}^*-\tilde{\bm{1}}_N=-\bm{r}_L\tilde{\bm{1}}_N\bm{r}_R^*.\label{eq:law2D_RT}
\ee
By taking the trace of both sides of this conservation law and slightly re-arranging the result, we find $\tilde{T} + \tilde{R}=\delta_{N,2M+1}$, or equivalently
\begin{align}
\re{\tilde{T}} + \re{\tilde{R}}&=
\begin{cases}
0,~N\;\;\text{even} \\
1,~N\;\;\text{odd}
\end{cases}
\label{eq:law2D2a_RT}\\
\im{\tilde{T}} + \im{\tilde{R}}&=0.\label{eq:law2D2b_RT}
\end{align}
Here $\tilde{T}\equiv Tr(\tilde{\bm{1}}_N\bm{t}^*\bm{t})$, $\tilde{R}\equiv Tr(\tilde{\bm{1}}_N\bm{r}_R^*\bm{r}_L)$ are complex in general, and $\delta_{N,2M+1}$ is the Kronecker delta where $M$ represents any positive integer. If the system has an even number of channels, then these relations imply in particular that $\tilde{T}$ and $\tilde{R}$ have the same modulus but are $\pi$ out of phase. We will refer to $\tilde{T}$ and $\tilde{R}$ as the generalized transmittance and reflectance in $\cal RT$-symmetric multimode waveguides.

\begin{figure}[t]
\begin{center}
\includegraphics[width=\linewidth]{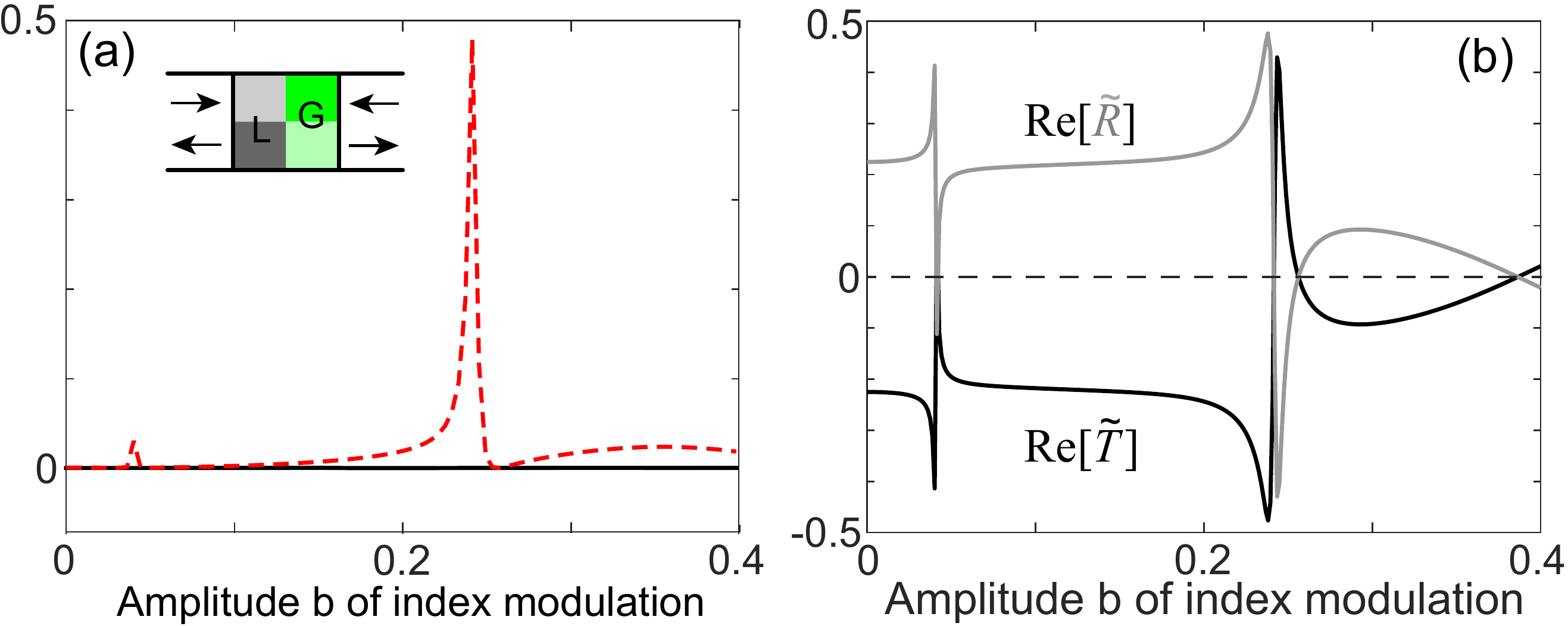}
\caption{Verification of generalized conservation laws in a 2D $\cal RT$-symmetric waveguide with two channels in both the left and right leads. (a) Solid line plots the norm of the difference between the two sides of Eq.~(\ref{eq:law2D_RT}), confirming the 2D conservation relation (\ref{eq:law2D_RT}). Dashed line plots the absolute value of the difference of the two sides of Eq.~(\ref{eq:law1D_RT}) for the $m=2$ channel, showing the breakdown of the 1D conservation relation (\ref{eq:law1D_RT}). (b) Black and grey lines plot $\re{\tilde{T}}$ and $\re{\tilde{R}}$. Their mirror symmetry about 0 (dashed horizontal line) verifies the 2D scalar conservation relation (\ref{eq:law2D2a_RT}), where $N=2$. The parameters are the same as in Fig.~\ref{fig:law2D}.}
\label{fig:law2D_RT}
\end{center}
\end{figure}

If the incident light is only in a single channel and it does not couple to other channels, Eq.~(\ref{eq:law2D_RT}) then reduces to its 1D form
\be
|T-1|=\sqrt{R_L R_R}, \label{eq:law1D_RT}
\ee
which is identical to that of the $\cal PT$-symmetric case with $T\equiv |\bm{t}_{mm}|^2$ and $R_{R,L}\equiv |(\bm{t}_{R,L})_{mm}|^2$. This 1D form is independent of $m$, i.e., whether the channel is an even or odd function of $y$, and more importantly, it does not require the system to be $\cal RT$- and $\cal PT$-symmetric simultaneously.

\begin{figure}[t]
\begin{center}
\includegraphics[width=\linewidth]{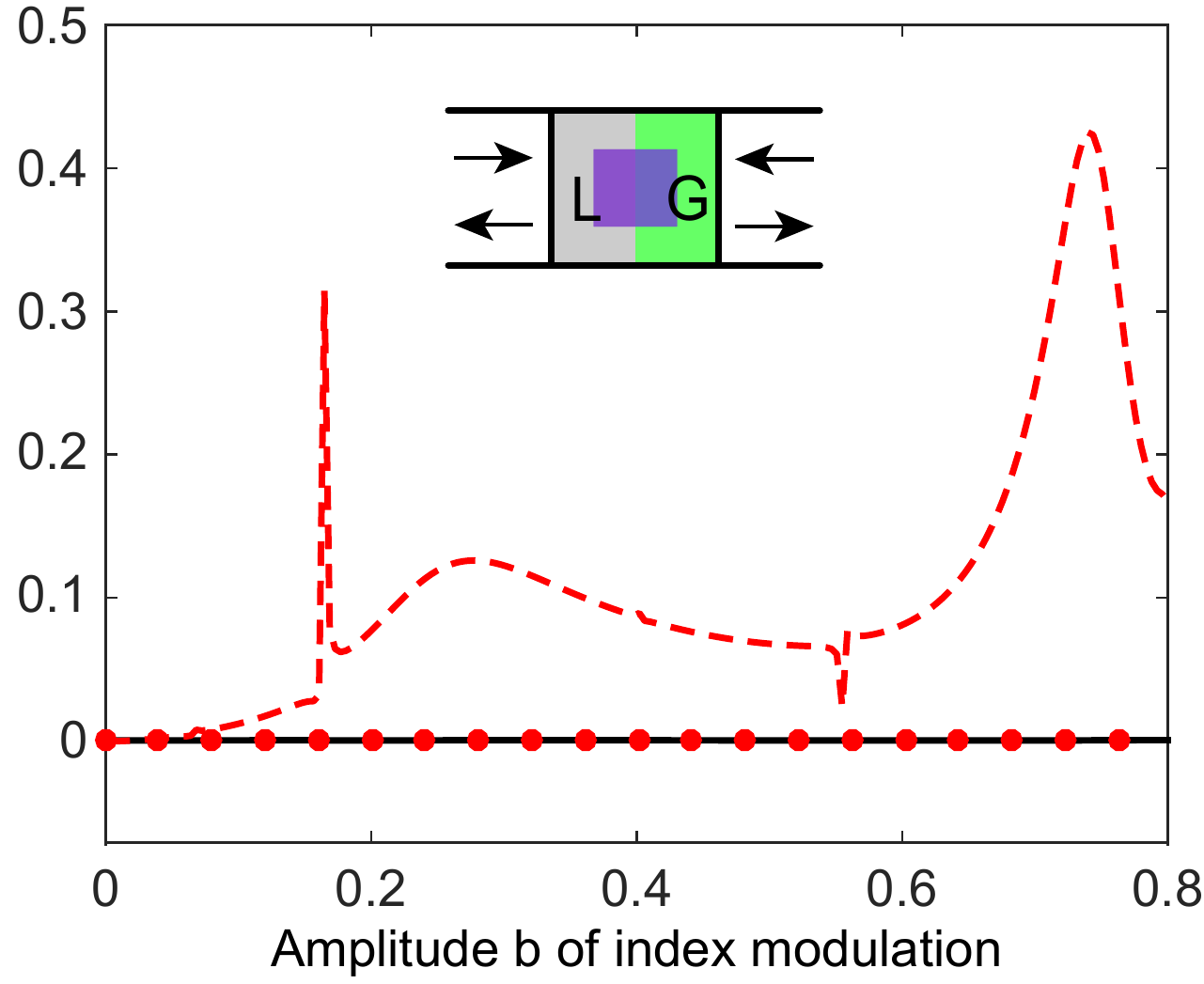}
\caption{Generalized conservation laws in a $\cal PT$- and $\cal RT$-symmetric waveguide that is non-separable in the longitudinal and transverse direction. Solid line shows the norm of the difference between the two sides of Eq.~(\ref{eq:law2D1}) (and that of Eq.~(\ref{eq:law2D_RT}) which is also zero). Dashed line and dots show the absolute value of the differences of the two sides of Eq.~(\ref{eq:law1D_RT}) for the $m=1$ and 2 channel respectively. The length $L$ and width $d$ of the waveguide are both 3 $\mu$m, and the index modulation is given by $\Delta n(x,y)=b \,(|x|,|y|<0.75 \mu{m})$. The other parameters are the same as in Fig.~\ref{fig:law2D}.}
\label{fig:PTRT}
\end{center}
\end{figure}

In Fig.~\ref{fig:law2D_RT} we again introduce a transverse index modulation $\Delta n$ to a half-gain-half-loss waveguide with a uniform $\re{n}$, similar to what we did in Fig.~\ref{fig:law2D} except that $\Delta n$ now depends on both $x$ and $y$ inside the waveguide, $\Delta n(x,y)=b\, (xy>0);\, -b\, (xy<0)$, which satisfies the $\cal RT$ symmetry but not the $\cal PT$-symmetry. At $b=0$ the system is separable and Eq.~(\ref{eq:law1D_RT}) holds for both channels. This is no longer the case as $b$ increases, and Eq.~(\ref{eq:law1D_RT}) breaks down while the generalized conservation law (\ref{eq:law2D_RT}) always holds.

We end this section by discussing a few properties of a multimode waveguide that is simultaneously $\cal RT$- and $\cal PT$-symmetric. In this case both Eq.~(\ref{eq:law2D1}) and (\ref{eq:law2D_RT}) hold, and by taking their difference we find that Eq.~(\ref{eq:law1D_RT}) also holds for each channel in a two-channel waveguide. It then indicates that there is no coupling between the channels, i.e., $\bm{t}$ and $\bm{r}_{L,R}$ are all diagonal, even though the system is not necessarily separable (see the 2D example in Fig.~\ref{fig:PTRT}). We note that this behavior only holds in the two-channel case: the system is symmetric about $y=0$, which is implied by the simultaneously satisfaction of $\cal RT$ and $\cal PT$ symmetries ($\cal RTPT=\cal RP$, i.e., $y\rightarrow -y$); the two channels are even and odd functions about $y=0$ respectively, and they cannot couple as a result. If there are more channels, then the even-$m$ channels couple to each other and so do the odd-$m$ channels. For example, we plot the difference of the left and right hand sides of Eq.~(\ref{eq:law1D_RT}) for the $m=1$ and 2 channels in a 3-channel waveguide in Fig.~\ref{fig:PTRT}(b). It is zero for the $m=2$ channel, which is the only channel of odd parity about $y=0$ and hence does not couple to the other two channels; it is nonzero for the $m=1$ channel, which couples to the other even-parity channel $m=3$. Since now the even-$m$ channels and the odd-$m$ channels form two uncoupled $\cal PT$-symmetric systems, we can define the generalized transmittance $\bar{T}$ and reflectance $\bar{R}$ in them respectively (i.e., $\bar{T}_{e,o}$ and $\bar{R}_{e,o}$), which are real and satisfy $\bar{T}_{e,o}+\bar{R}_{e,o}=N_{e,o}$ according to Eq.~(\ref{eq:law2D3}), where $N_{e,o}$ are the number of even-$m$ and odd-$m$ channels. This observation then implies that the other generalized transmittance $\tilde{T}$ and reflectance $\tilde{R}$ for the whole ($\cal RT$-symmetric) system are now also real valued and given by
\begin{align}
\tilde{T}&=\bar{T}_o-\bar{T}_e, \\
\tilde{R}&=\bar{R}_o-\bar{R}_e,
\end{align}
which is the simplest scenario that leads to the conservation relation (\ref{eq:law2D2a_RT}). We emphasize that these two relations above hold only when the system is simultaneously $\cal RT$- and $\cal PT$-symmetric.

\section{Spontaneous Symmetry Breaking of the $\bm{S}$ Matrix}

\begin{figure}[b]
\begin{center}
\includegraphics[width=\linewidth]{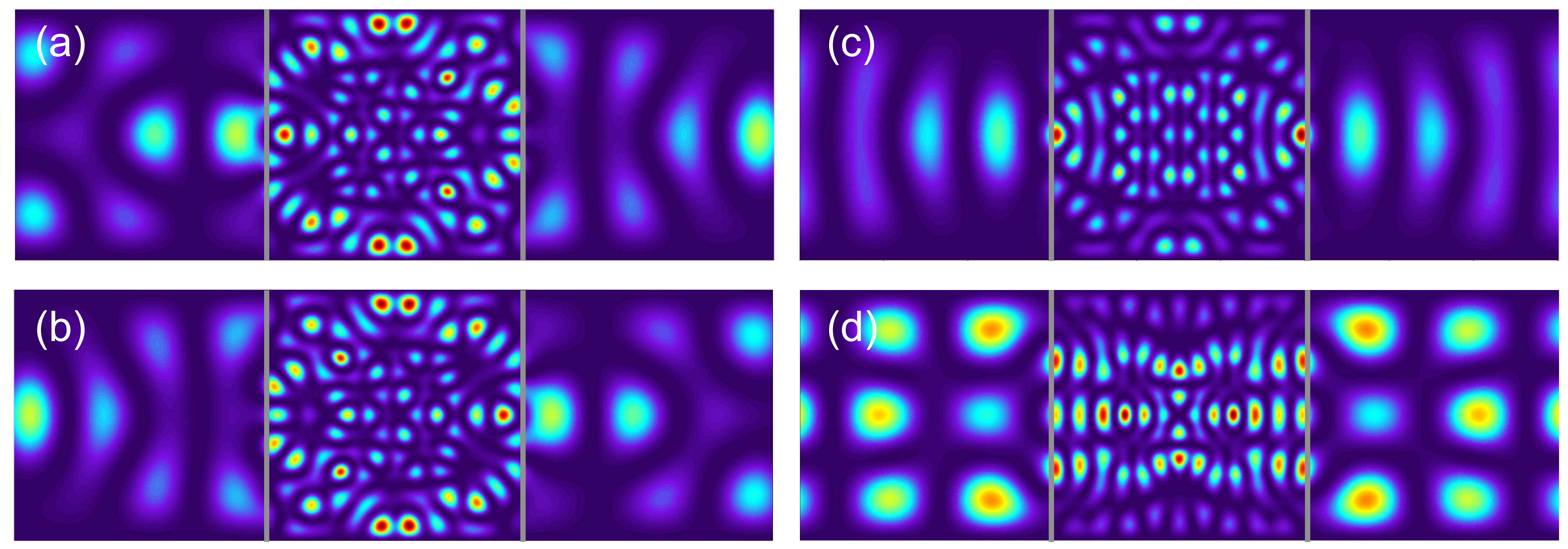}
\caption{ (Color online) Intensity plot of scattering eigenstates in the $\cal PT$- and $\cal RT$-symmetric multimode waveguide shown in Fig.~\ref{fig:PTRT}. (a,b) A pair of $\cal PT$- and $\cal RT$-broken eigenstates. (c,d) A pair in the $\cal PT$- and $\cal RT$-symmetric phase. All four eigenstates are the result of the coupling between $m=1,3$ channels. $\im{n}=\pm0.05$, $b=0.8$, and the grey vertical lines mark the left and right sides of the waveguide.}
\label{fig:wf}
\end{center}
\end{figure}

Due to the symmetry relation (\ref{eq:RTSRT}) in a $\cal RT$-symmetric system, the $S$ matrix can undergo a spontaneous symmetry breaking, similar to the $\cal PT$-symmetric case. If the system is simultaneously $\cal PT$- and $\cal RT$-symmetric, one may wonder whether one of these two symmetries can be broken while the other one is not. This situation does not occur for two reasons. First, if one symmetry is broken then some or all the eigenvalues of the $S$ matrix are no longer unimodular, but if the other symmetry still holds, then all the eigenvalues should be unimodular, which imposes an obvious contradiction. The same argument applies to a specific pair of scattering eigenstates $\bm{\varphi}_1,\bm{\varphi}_2$, i.e., they cannot be $\cal PT$-broken but still $\cal RT$-symmetric (or vice versa). Another way to arrive at this observation is the following. If $\bm{\varphi}_1,\bm{\varphi}_2$ are in the $\cal PT$-broken phase, then ${\cal PT}\bm{\varphi}_1\propto\bm{\varphi}_2$ holds \cite{CPALaser}. As we mentioned at the end of the previous section, the system itself has parity symmetry about $y=0$ when it is both $\cal PT$- and $\cal RT$-symmetric, meaning that $\bm{\varphi}_1$ and $\bm{\varphi}_2$ are either even or odd functions of $y$. We then find
\be
{\cal RT}\bm{\varphi}_1 \propto \bm{\varphi}_2,
\ee
using ${\cal P}_y{\cal PT} = {\cal RT}$, which indicates that $\bm{\varphi}_1$ and $\bm{\varphi}_2$ are also in the $\cal RT$-broken phase; otherwise we would find ${\cal RT}\bm{\varphi}_{1,2} \propto \bm{\varphi}_{1,2}$ instead. Here ${\cal P}_y$ is the parity operator about $y=0$. Since this derivation is reversible, it gives the second reason why the two symmetries either hold or break simultaneously. As an example, we show in Figs.~\ref{fig:wf}(a) and (b) such a pair of broken-symmetry scattering eigenstates, the intensity profile of which can be easily checked to satisfy ${\cal P}|\bm{\varphi}_1|^2 \propto |\bm{\varphi}_2|^2$ and ${\cal R}|\bm{\varphi}_1|^2 \propto|\bm{\varphi}_2|^2$, as a consequence of ${\cal PT}\bm{\varphi}_1\propto\bm{\varphi}_2$ and ${\cal RT}\bm{\varphi}_1\propto\bm{\varphi}_2$, respectively. Figs.~\ref{fig:wf}(c) and (d) show a pair of scattering eigenstates that are both $\cal PT$- and $\cal RT$-symmetric, the intensity of which satisfies both ${\cal P}|\bm{\varphi}_{1,2}|^2 \propto |\bm{\varphi}_{1,2}|^2$ and ${\cal R}|\bm{\varphi}_{1,2}|^2 \propto |\bm{\varphi}_{1,2}|^2$.

The simplest multimode waveguide that is simultaneously $\cal PT$- and $\cal RT$-symmetric is a separable half-gain-half-loss system depicted in Fig.~\ref{fig:schematic}(a). Below we discuss how spontaneous symmetry breaking of the $S$ matrix can be observed in such a multimode waveguide, \textit{without} the need to tune any system parameters. It is based on the observation that scattering in uncoupled channels of different transverse order $m$, at a given frequency, displays different thresholds for symmetry breaking in terms of the gain and loss strength. Therefore, if a system possesses $\cal PT$-symmetry at a particular frequency $\omega_0$, the scattering at this frequency will display two contrasting behaviors depending on the scattering channel: it can either be in the $\cal PT$ symmetric phase with conserved flux in the corresponding scattering eigenstates, or in the broken symmetry phase with a pair of amplified and attenuated scattering eigenstates. The transverse order $m$ then plays the role of the symmetry breaking parameter.

We have chosen the waveguide with uncoupled channels for a practical consideration: scattering eigenstates that exist in more than one channel is difficult to measure and tune in an experiment. Since we want to observe a clear transition between the symmetric and broken phases of the $S$ matrix, we need at least a few channels in both phases. %Hence we will not consider the non-separable two-channel $\cal RT$- and $\cal PT$-symmetric system discussed at the end of the previous section, even thought its two channels are also decoupled.

The $m$-dependent symmetry breaking can be understood in two ways. First we note that the symmetry breaking threshold (and the EP) of the $S$ matrix for a 1D half-loss-half-gain structure is given by the following expression \cite{CPALaser}:
\be
\frac{\omega_c}{c}L \approx \frac{1}{\tau}\ln\left(\frac{2n_0}{\tau}\right), \label{eq:TH}
\ee
where $n_0\pm i\tau$ is the complex refractive index in the loss and gain regions and $c$ is the speed of light in vacuum. For a given $n_0$ and $\tau$, the broken symmetry phase lies in $\omega_0>\omega_c$ and the symmetric phase in $\omega_0<\omega_c$. In the semiclassical regime where the wavelength is much shorter than the system size, one can apply the picture of ray optics, and propagating modes of different transverse order $m$ experience a different length $L\rightarrow L_m$ in the scattering region, because they propagate at different angles with respective to the sidewalls (e.g., see the wave vectors $k_x,k_y$ in Fig.~\ref{fig:schematic}(b)). As a result, Eq.~(\ref{eq:TH}) shows that $\omega_c$ at the symmetry breaking threshold is now $m$-dependent (denoted by $\omega_c^{(m)}$) for a given $n_0$ and $\tau$. Consequently, scattering in higher transverse channels (with a larger $m$, a longer $L_m$, and a lower $\omega_c^{(m)}$) can be in the broken symmetry phase ($\omega_0>\omega_c^{(m)}$), while scattering in lower transverse channels can be in the $\cal PT$-symmetric phase ($\omega_0<\omega_c^{(m)}$) at the same frequency $\omega_0$.

\begin{figure}[t]
\begin{center}
\includegraphics[width=\linewidth]{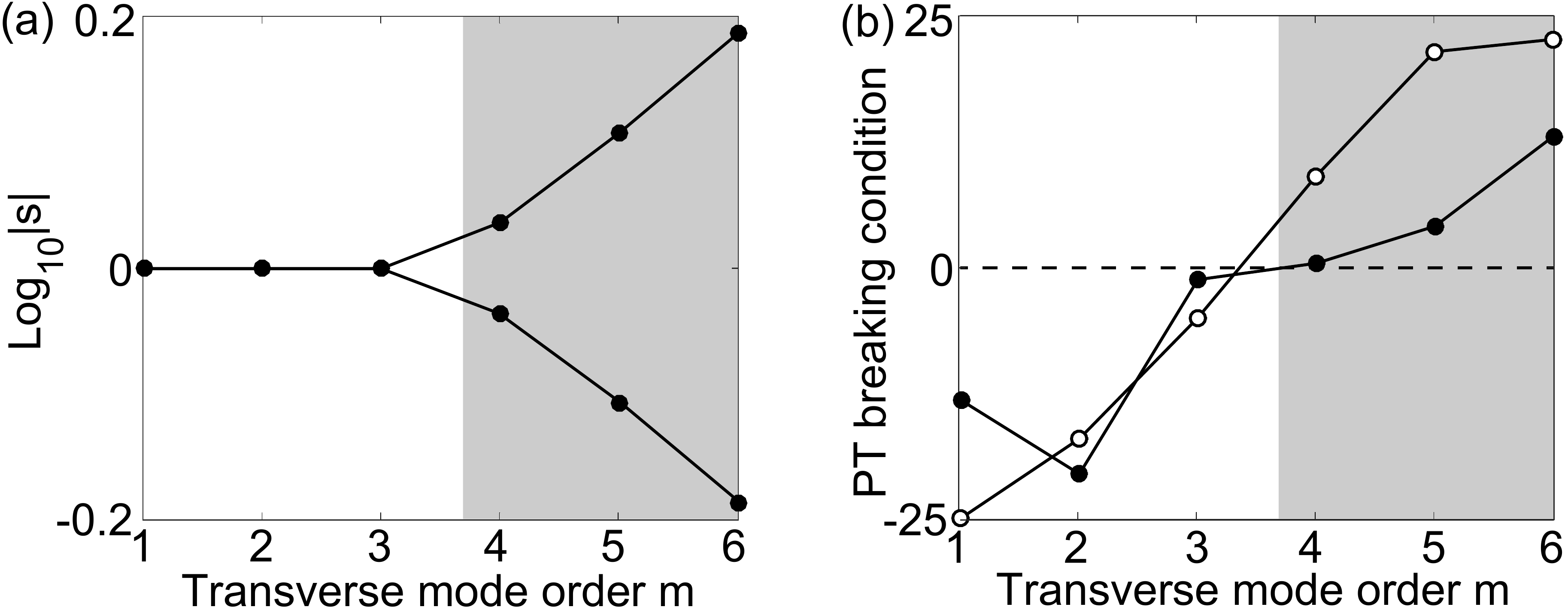}
\caption{(a) Spontaneous symmetry breaking of the eigenvalues $s$ of the scattering matrix in a multimode waveguide. The parameters used are $n_0=3,\tau=0.005,L=23\,\mu{m},d=5\,\mu{m}$, and the wavelength is $1550\,nm$ ($\omega_0/c=4.05\,{\mu}m^{-1}$). At this frequency the waveguide supports six channels ($m=1,2,\ldots,6$), and the upper three are in the breaking symmetry phase (shaded area). (b) Filled circles: the difference of the left and right hand sides of the symmetry breaking condition (\ref{eq:TH2}), which predicts the same broken symmetry regime ($m\geq4$) as in (a). Open circles: 10 times the value of $(R_L+R_R)-2(T+1)$ in each channel, the sign of which indicates whether the system is in the $\cal PT$-symmetric phase or broken symmetry phase.
%(c,d) False color intensity plots of the scattering eigenstates of $m=1$ (in the $\cal PT$-symmetric phase) and 5 (in the broken symmetry phase), respectively. White shaded area represents the waveguide.
}
\label{fig:transition}
\end{center}
\end{figure}

To be more quantitative and able to address the regime where the system size is comparable to the wavelength, we offer an alternative (but equivalent) explanation of the channel-dependent symmetry breaking, by mapping the wave propagation to an effective 1D system where the criterion (\ref{eq:TH}) holds. As we shall see, in this explanation the length $L$ of the scattering region is the same in all channels, while the effective frequency and refractive index now become channel-dependent.
Take the 2D waveguide shown in Fig.~\ref{fig:schematic} for example, the transverse wave number is $k_y=m\pi/d,\,(m=1,2,\ldots)$, and the incident light thus has an effective frequency
\be
\omega_m = \sqrt{\omega_0^2-c^2k_y^2}
\ee
in the propagation direction $x$, and the effective index inside the gain and loss regions is
\be
n_m = \frac{\sqrt{(n_0\pm i\tau)^2\omega_0^2-c^2k_y^2}}{\omega_m},
\ee
which is still $\cal PT$ symmetric. The criterion (\ref{eq:TH}) now becomes
\be
\frac{\omega_m}{c}L \approx \frac{1}{|\im{n_m}|}\ln\left(\frac{2\re{n_m}}{|\im{n_m}|}\right), \label{eq:TH2}
\ee
and it predicts a symmetry breaking threshold between $m=3$ and 4 for the case shown in Fig.~\ref{fig:transition}, which is verified by directly calculating the eigenvalues of the $S$ matrix.

In Ref.~\cite{conservation} a simplified signature of the symmetry breaking was given in terms of the transmittance and reflectances for a 1D $\cal PT$-symmetric structure: the system is in the symmetric phase if one measures $(R_L+R_R)-2(T-1)<0$; otherwise it is in the broken symmetry phase. We test this signature in each channel using $T\equiv |\bm{t}_{mm}|^2$ and $R_{L,R}\equiv |(\bm{r}_{L,R})_{mm}|^2$ in the example above, and indeed this quantity changes its sign between $m=3$ and 4 (see the open circles in Fig.~\ref{fig:transition}(b)), which makes the detection of the symmetry breaking much easier than tuning into the scattering eigenstates and measuring their scattering eigenvalues. The latter can, of course, be measured indirectly by substituting the measured values of $\bm{t}_{mm},(\bm{r}_{L,R})_{mm}$ in the definition of the $S$ matrix (Eq.~(\ref{eq:S2D}) for each channel), which has the inconvenience of measuring the phases of $\bm{t}_{mm},(\bm{r}_{L,R})_{mm}$.

\section{Conclusion}

In summary, we have discussed the generalized conservation laws of scattering in $\cal PT$-symmetric and $\cal RT$-symmetric multimode waveguides. Not only do they exist in a matrix form for the transmission and reflection matrices, they also exist in a scalar form for real-valued quantities using generalized transmittances and reflectances. If different waveguide channels are decoupled, each channel is effectively a 1D system and we recover the generalized conservation law found in Ref.~\cite{conservation}, now for both $\cal PT$-symmetric and $\cal RT$-symmetric cases. 
%The decoupling of waveguide channels does not necessarily require the system to be separable in the longitudinal and transverse direction. 
When this condition holds, we have shown that a simple half-gain-half-loss $\cal PT$-symmetric system offers a convenient setup to observe the spontaneous symmetry breaking of the $S$ matrix, without the need to tune any system parameters. It is facilitated by the existence of a single ``exceptional" value of the channel number $m$ for a given incident light frequency and fixed gain and loss strength in this system. By performing scattering experiment in each channel of the waveguide, one observes the transition from the symmetric phase to the broken symmetry phase as $m$ becomes larger than this exceptional value.

\section*{Acknowledgement}

We thank Douglas Stone for helpful discussions. L.G. acknowledges support by NSF under grant No. DMR-1506987. L.F. acknowledges support by NSF under grant No. DMR-1506884 and by US Army Research Office under grant No. W911NF-15-1-0152. K.G.M. is supported by the People Programme (Marie Curie Actions) of the European Union's Seventh Framework Programme (FP7/2007-2013) under REA grant agreement number PIOF-GA-2011- 303228 (project NOLACOME) and by the European Union Seventh Framework Programme (FP7-REGPOT-2012-2013-1) under grant agreement 316165.

\appendix
\section*{Appendix: Eigenvalues of $\bm{A}\bm{A}^*$}
Interestingly, we find that the eigenvalues of the matrix $\bm{A}\bm{A}^*$ are either real or form complex conjugate pairs, where $\bm{A}$ is an arbitrary square matrix. Here we provide a short proof. If $\bm{A}\bm{A}^*$ has an eigenvalue $\lambda_n$ and the corresponding eigenvector is $\bm{\psi}_n$, i.e.,
\be
\bm{A}\bm{A}^*\bm{\psi}_n=\lambda_n\bm{\psi}_n,
\ee
then by taking the complex conjugate of both sides and multiplying $\bm{A}$ from left, we find
\be
\bm{A}\bm{A}^*\bm{A}\bm{\psi}_n^*=\lambda_n^*\bm{A}\bm{\psi}_n^*.
\ee
It indicates that $\lambda_n^*$ is also an eigenvalue of $\bm{A}\bm{A}^*$, with the corresponding eigenvector $\bm{A}\bm{\psi}_n^*$. Therefore, $\lambda_n$ is real if $\bm{A}\bm{\psi}_n^*\propto\bm{\psi}_n$, or a pair of complex conjugate eigenvalues $\lambda_n,\lambda_{n'}$ exist when $\bm{A}\bm{\psi}_n^*\propto\bm{\psi}_{n'}\,(n\neq n')$.

This property not only guarantees that $Tr(\bm{A}\bm{A}^*)$ (and more specifically $Tr(\bm{t}\bm{t}^*)$) is real, which we have used in the main text to derive the real-valued conservation law (\ref{eq:law2D3}); it also suggests a general form of the effective Hamiltonian for a $\cal PT$-symmetric (and $\cal RT$-symmetric) system. For example, the toy model of two coupled $\cal PT$-symmetric resonators can be written as
\be
\bm{H} =
\begin{pmatrix}
\omega+i\tau & g \\
g & \omega - i\tau
\end{pmatrix},
\ee
where $\omega$ is the identical resonant frequency of both resonators, $\pm\tau$ represent the loss and gain strength, and $g\ll\omega$ is the coupling strength of these two resonators. The decomposition $\bm{H}=\bm{A}\bm{A}^*$ exists for multiple choices of $\bm{A}$, and one such choice is
\be
\bm{A} =
\begin{pmatrix}
\frac{g}{2b} & b \\
b - i\frac{\tau}{b} & \frac{g}{2b}
\end{pmatrix},
\ee
where $b$ is a real quantity satisfying $b^2 = (\omega\pm\sqrt{\omega^2-g^2})/2$.

\bibliographystyle{longbibliography}

\begin{thebibliography}{100}
\bibitem{El-Ganainy_OL06} R.~El-Ganainy, K.~G.~Makris, D.~N.~Christodoulides, and Z.~H.~Musslimani, \textit{Theory of Coupled Optical $\cal PT$-Symmetric Structures}, Opt. Lett. {\bf 32}, 2632 (2007).
\bibitem{Moiseyev} S.~Klaiman, U.~Gunther, and N.~Moiseyev, \textit{Visualization of Branch Points in $\cal PT$-Symmetric Waveguides}, Phys. Rev. Lett. {\bf 101}, 080402 (2008).
\bibitem{Musslimani_prl08} Z.~H.~Musslimani, K.~G.~Makris, R.~El-Ganainy, D.~N.~Christodoulides, \textit{Optical Solitons in $\cal PT$ Periodic Potentials}, Phys. Rev. Lett. {\bf 100}, 030402 (2008).
\bibitem{Makris_prl08} K.~G.~Makris, R.~El-Ganainy, D.~N.~Christodoulides, and Z.~H.~Musslimani, \textit{Beam Dynamics in $\cal PT$ Symmetric Optical Lattices}, Phys. Rev. Lett. {\bf 100}, 103904 (2008).
\bibitem{Guo} A. Guo, G. J. Salamo, D. Duchesne, R. Morandotti, M. Volatier-Ravat, V. Aimez, G. A. Siviloglou, and D. N. Christodoulides, \textit{Observation of PT-Symmetry Breaking in Complex Optical Potentials}, Phys. Rev. Lett. \textbf{103}, 093902 (2009).
\bibitem{mostafazadeh} A.~Mostafazadeh, \textit{Spectral Singularities of Complex Scattering Potentials and Infinite Reflection and Transmission Coefficients at Real Energies}, Phys. Rev. Lett. {\bf 102}, 220402 (2009).
\bibitem{Kottos} T.~Kottos, \textit{Optical Physics: Broken Symmetry Makes light Work}, Nat. Phys. {\bf 6}, 166 (2010).
\bibitem{Makris_PRA10} K. G. Makris, R. El-Ganainy, D. N. Christodoulides, and Z. Musslimani, \textit{PT Symmetric Optical Lattices}, Phys. Rev. A
\textbf{81}, 063807 (2010).
\bibitem{CPALaser} Y.~D.~Chong, L.~Ge, and A.~D.~Stone, \textit{$\cal PT$-Symmetry Breaking and Laser-Absorber Modes in Optical Scattering systems}, Phys. Rev. Lett. {\bf 106}, 093902 (2011).
\bibitem{Longhi} S. Longhi, \textit{$\cal PT$-Symmetric Laser Absorber}, Phys. Rev. A \textbf{82}, 031801(R) (2010).
\bibitem{conservation} L.~Ge, Y.~D.~Chong, and A. D. Stone, \textit{Conservation Relations and Anisotropic Transmission Resonances in One-Dimensional $\cal PT$-Symmetric Photonic Heterostructures}, Phys. Rev. A {\bf 85}, 023802 (2012).
\bibitem{Robin} P. Ambichl, K. G. Makris, L.Ge, Y. Chong, A. D. Stone, and S. Rotter, 
\textit{Breaking of PT Symmetry in Bounded and Unbounded Scattering Systems}, 
Phys. Rev. X {\bf 3}, 041030 (2013).
\bibitem{EP9} L. Ge and A. D. Stone, \textit{Parity-time symmetry breaking beyond one dimension: the role of degeneracy},
Phys. Rev. X \textbf{4}, 031011 (2014).
\bibitem{Lin} Z. Lin, H. Ramezani, T. Eichelkraut, T. Kottos, H. Cao, and D. N. Christodoulides, \textit{Unidirectional Invisibility Induced
by PT-Symmetric Periodic Structures}, Phys. Rev. Lett. \textbf{106}, 213901 (2011).
\bibitem{Ruter} C. E. R\"uter, K. G. Makris, R. El-Ganainy, D. N. Christodoulides, M. Segev, and D. Kip, \textit{Observation of Parity-Time Symmetry in Optics}, Nat. Phys. \textbf{6}, 192 (2010).
\bibitem{Feng} L. Feng, M. Ayache, J. Huang, Y.-L. Xu, M.-H. Lu, Y.-F. Chen, Y. Fainman, and A. Scherer, \textit{Nonreciprocal Light Propagation in a Silicon Photonic Circuit}, Science \textbf{333}, 729 (2011).
\bibitem{Feng_NM} L. Feng, Y.-L. Xu, W. S. Fegadolli, M.-H. Lu, J. E. B. Oliveira, V. R. Almeida, Y.-F. Chen, and A. Scherer, \textit{Experimental Demonstration of a Unidirectional Reflectionless Parity-Time Metamaterial at Optical Frequencies}, Nat. Mater. \textbf{12}, 108 (2013).
\bibitem{Feng2}L. Feng, Z. J.Wong, R.-M.Ma, Y.Wang, and X. Zhang, \textit{Singlemode laser by parity-time symmetry breaking}, Science \textbf{346}, 972
(2014).
\bibitem{Walk} A. Regensburger, C. Bersch, M. A. Miri, G. Onishchukov, D. N. Christodoulides, and U. Peschel, \textit{Parity-Time Synthetic Photonic Lattices}, Nature (London) \textbf{488}, 167 (2012).
\bibitem{Hodaei} H. Hodaei, M. A. Miri, M. Heinrich, D. N. Christodoulides, and M. Khajavikhan, \textit{Parity-time–symmetric microring lasers},
Science \textbf{346}, 975 (2014).
\bibitem{Yang} B. Peng, S. K. \"Ozdemir, F. Lei, F. Monifi, M. Gianfreda, G. L. Long, S. Fan, F. Nori, C. M. Bender, and L. Yang, \textit{Parity-Time-Symmetric Whispering-Gallery Microcavities}, Nat Phys, \textit{10}, 394 (2014).

\bibitem{EP1} J. Okolowicz, M. Ploszajczak, and I. Rotter, \textit{Dynamics of Quantum Systems Embedded in a Continuum}, Phys. Rep. {\bf 374}, 271 (2003).
\bibitem{EP2} W. D. Heiss, \textit{Exceptional Points of Non-Hermitian Operators}, J. Phys. A: Math. Gen. {\bf 37}, 2455 (2004).
\bibitem{EPMVB} M. V. Berry, \textit{Physics of Nonhermitian Degeneracies}, Czechoslovak J. Phys. {\bf 54}, 1039 (2004).
\bibitem{EP3} N. Moiseyev, {\it Non-Hermitian Quantum Mechanics} (Cambridge, New York, 2011).
\bibitem{EP4} C. Dembowski, H.-D. Gr\"af, H. L. Harney, A. Heine, W. D. Heiss, H. Rehfeld, and A. Richter, \textit{Experimental Observation of the Topological Structure of Exceptional Points}, Phys. Rev. Lett. {\bf 86}, 787 (2001).
\bibitem{EP5} J. Wiersig, S.-W. Kim, and M. Hentschel, \textit{Asymmetric Scattering and Nonorthogonal Mode Patterns in Optical Microspirals}, Phys. Rev. A {\bf 78}, 053809 (2008).
\bibitem{EP6} S.-B. Lee, J. Yang, S. Moon, S.-Y. Lee, J.-B. Shim, S. W. Kim, J.-H. Lee, and K. An, \textit{Observation of an Exceptional Point in a Chaotic Optical Microcavity}, Phys. Rev. Lett. {\bf 103}, 134101 (2009).
\bibitem{EP7} A.~Guo, G.~J.~Salamo, D.~Duchesne, R.~Morandotti, M.~Volatier-Ravat, V.~Aimez, G.~A.~Siviloglou and D.~N.~Christodoulides, \textit{Observation of PT-Symmetry Breaking in Complex Optical Potentials},
    Phys. Rev. Lett. {\bf 103}, 093902 (2009).
\bibitem{EP8} M.~Liertzer, L.~Ge, A.~Cerjan, A.~D.~Stone, H.~E.~T\"{u}reci, and S.~Rotter, \textit{Pump-Induced Exceptional Points in Lasers}, Phys. Rev. Lett. {\bf 108}, 173901 (2012).
\bibitem{Bender1} C.~M.~Bender and S.~Boettcher, \textit{Real Spectra in Non-Hermitian Hamiltonians Having $\cal PT$ Symmetry}, Phys. Rev. Lett. {\bf 80}, 5243 (1998).
\bibitem{Bender2} C.~M.~Bender, S.~Boettcher, and P.~N.~Meisinger, \textit{$\cal PT$-Symmetric Quantum Mechanics}, J. Math. Phys. {\bf 40}, 2201 (1999).
\bibitem{Bender3} C.~M.~Bender, D.~C.~Brody, and H.~F.~Jones, \textit{Complex Extension of Quantum Mechanics}, Phys. Rev. Lett. {\bf 89}, 270401 (2002).
\bibitem{Haus} H. A. Haus, \textit{Waves and Fields in Optoelectronics} (Prentice-Hall, Englewood Cliffs, NJ, 1984), pp. 56-–61.
\bibitem{Collin} R. E. Collin, \textit{Field Theory of Guided Waves} (McGraw-Hill, New York, 1960).
\bibitem{Landau} L. D. Landau and E. M. Lifshitz, \textit{Electrodynamics of Continuous Media} (Pergamon Press, Oxford, 1960).
\bibitem{Pagneux} V. Pagneux et al.,  private communication.
\bibitem{Beenakker} C. w. Beenakker, \textit{Random-Matrix Theory of quantum transport}, Rev. Mod. Phys. \textbf{69}, 731--808 (1997).
\end{thebibliography}

\end{document}